\documentclass[apj]{emulateapj}

\slugcomment{accepted for publication in ApJ}
\shorttitle{A first site of galaxy cluster formation}
\shortauthors{Toshikawa et al.}

\begin{document}

\title{A First Site of Galaxy Cluster Formation: Complete Spectroscopy of a Protocluster at $z=6.01$}
\author{Jun Toshikawa\altaffilmark{1,2}, Nobunari Kashikawa\altaffilmark{1,2}, Roderik Overzier\altaffilmark{3},
    Takatoshi Shibuya\altaffilmark{4}, Shogo Ishikawa\altaffilmark{1,2}, Kazuaki Ota\altaffilmark{5,6},
    Kazuhiro Shimasaku\altaffilmark{7}, Masayuki Tanaka\altaffilmark{2}, Masao Hayashi\altaffilmark{2},
    Yuu Niino\altaffilmark{2}, Masafusa Onoue\altaffilmark{1,2}}
\email{jun.toshikawa@nao.ac.jp}
\altaffiltext{1}{Department of Astronomy, School of Science, Graduate University for Advanced Studies, 
    Mitaka, Tokyo 181-8588, Japan}
\altaffiltext{2}{Optical and Infrared Astronomy Division, National Astronomical Observatory, 
    Mitaka, Tokyo 181-8588, Japan}
\altaffiltext{3}{Observat\'{o}rio Nacional, Rua Jos\'{e} Cristino, 77. CEP 20921-400, S\~{a}o Crist\'{o}v\~{a}o,
    Rio de Janeiro-RJ, Brazil}
\altaffiltext{4}{Institute for Cosmic Ray Research, The University of Tokyo, 5-1-5 Kashiwanoha, Kashiwa, Chiba
    277-8582, Japan}
\altaffiltext{5}{Kavli Institute for Cosmology, University of Cambridge, Madingley Road, Cambridge CB3 0HA, UK}
\altaffiltext{6}{Cavendish Laboratory, University of Cambridge, 19 J.J. Thomson Avenue, Cambridge CB3 0HE, UK}
\altaffiltext{7}{Department of Astronomy, University of Tokyo, Hongo, Tokyo 113-0033, Japan.}

\begin{abstract}
We performed a systematic spectroscopic observation of a protocluster at $z=6.01$ in the Subaru Deep Field.
We took spectroscopy for all 53 $i'$-dropout galaxies down to $z'=27.09\,\mathrm{mag}$ in/around the
protocluster region.
From these observations, we confirmed that 28 galaxies are at $z\sim6$, of which ten are clustered in a narrow
redshift range of $\Delta z<0.06$.
To trace the evolution of this primordial structure, we applied the same $i'$-dropout selection and the same
overdensity measurements used in the observations to a semi-analytic model built upon the Millennium Simulation.
We obtain a relation between the significance of overdensities observed at $z\sim6$ and the predicted dark
matter halo mass at $z=0$.
This protocluster with $6\sigma$ overdensity is expected to grow into a galaxy cluster with a mass of
$\sim5\times10^{14}\,\mathrm{M_\sun}$ at $z=0$.
Ten galaxies within $10\,\mathrm{comoving\>Mpc}$ of the overdense region can, with more than an 80\%
probability, merge into a single dark matter halo by $z=0$.
No significant differences appeared in UV and Ly$\alpha$ luminosities between the protocluster and field galaxies,
suggesting that this protocluster is still in the early phase of cluster formation before the onset of any obvious
environmental effects.
However, further observations are required to study other properties, such as stellar mass, dust, and age.
We do find that galaxies tend to be in close pairs in this protocluster.
These pair-like subgroups will coalesce into a single halo and grow into a more massive structure.
We may witness an onset of cluster formation at $z\sim6$ toward a cluster as seen in local universe.
\end{abstract}


\section{INTRODUCTION}
Revealing structure formation and galaxy evolution are compelling objectives in modern astronomy.
In the local universe, many studies have shown that these two topics are closely related.
Star-formation activity are quenched in high density environments, such as galaxy clusters, where red and
early-type galaxies dominate.
These galaxies represent the characteristic features of galaxy clusters, such as red sequences or
morphology-density relation \citep{visvanathan77,dressler80}.
Even in modest galaxy-groups environments, the star formation is known to be effectively quenched at $z<0.1$
\citep[e.g.,][]{rasmussen12}.
Especially, massive and bright elliptical galaxies in clusters have significantly different properties from their
field counterparts, such as higher stellar velocity dispersion and higher $\alpha$/Fe ratio.
These differences suggest that elliptical galaxies in galaxy clusters contain more dark matter and involve a
shorter star-formation timescale \citep{thomas05,linden07}.
These are intuitively expected within the hierarchical structure formation scenario: halos in higher-density regions
should collapse earlier and merge more rapidly, which causes earlier galaxy formation and more rapid evolution
\citep{kauffmann99,benson01,springel05,lucia06}.
In this way, both observational and theoretical studies predict that galaxy evolution strongly depends on the
environment.
Therefore, direct observation of protoclusters, which are high density regions at high redshift, is required to
clarify the relation between galaxy evolution and structure formation over cosmic time.

The fraction of star-forming galaxies in a galaxy cluster increases at higher redshift
\citep{butcher84,haines09,lerchster11}, and a higher star-formation rate (SFR) is observed in higher-density
environments at $z\sim1$ \citep[e.g.,][]{popesso11}.
At $z>2$, massive clusters as seen in the local universe have not yet formed; protoclusters are growing by merging
and activating the star formation through accretion of material from their surroundings. 
Thus, protoclusters are identified as overdense regions of star-forming galaxies such as Lyman break galaxies
(LBGs) and Ly$\alpha$ emitters (LAEs) \citep{malkan96,steidel98,venemans07,mawatari12,lee14}.
Although star-forming and young galaxies are the majority, many red and massive galaxies certainly exist in
some protoclusters, suggesting that environmental effects are actually seen at least up to $z\sim2-3$
\citep{kodama07,zirm08,kubo13,lemaux14}.
Protocluster galaxies have higher stellar mass than their field counterparts \citep{steidel05,kuiper10,hatch11},
and exceptional objects, such as Ly$\alpha$ blobs, submillimeter galaxies, and active galactic nuclei, are likely
to be discovered in high density environments \citep{digby10,tamura10,matsuda11}.
What causes this difference between protocluster regions and field regions?
In order to address this question, further observations of higher redshift protoclusters are important to probe
the onset of initial environmental effects in the early universe.
The findings will provide important information about the relation between the galaxy and environment during the
first stage of cluster formation.
Although the physical properties of protoclusters are not fully revealed at higher redshifts due to the
difficulties involved in multi-wavelength imaging and the small number of spectroscopic confirmations, some
studies pushed forward searching for high-redshift protoclusters.
A handful of samples of protoclusters was discovered at $z>4$ \citep{ouchi05,venemans07,overzier09,capak11}, and
some overdense regions without spectroscopic confirmation were identified at even higher redshifts
\citep{malhotra05,zheng06,stiavelli05,trenti12}.

In this paper, we extend the previous spectroscopic follow-up observations presented in \citet{toshikawa12}.
We have carried out a further spectroscopic observations of a protocluster at $z=6.01$ in the Subaru Deep Field
(SDF), which is the highest redshift protocluster known to date with sufficient spectroscopic confirmation.
Based on these complete spectroscopic observations, we attempt to investigate whether there are any differences
between the protocluster and field galaxies at $z=6$ by analyzing UV/Ly$\alpha$ luminosities and the
three-dimensional distribution of protocluster galaxies.

The structure of this paper is as follows: we summarize the previous and new observations in Section \ref{obs} and
describe the redshift identifications in Section \ref{result}.
Section \ref{model} compares the results with model predictions.
We discuss the properties and the structure of the protocluster in Section \ref{discuss}.
The conclusions are presented in Section \ref{conc}.
We assume the following cosmological parameters: $\Omega_\mathrm{M}=0.3, \Omega_\Lambda=0.7, 
\mathrm{H}_0=70 \mathrm{\,km\,s^{-1}\,Mpc^{-1}}$, which yield an age of the universe of $910\,\mathrm{Myr}$
and a spatial scale of $40\,\mathrm{kpc\,arcsec^{-1}}$ in comoving units at $z=6.01$.
Unless otherwise noted, we use comoving units throughout, and magnitudes are given in the AB system.

\section{OBSERVATIONS} \label{obs}
\subsection{A Protocluster at $z=6.01$}
In our previous paper \citep{toshikawa12}, we reported the discovery of a protocluster at $z=6.01$.
The details of the analysis, such as sample selection, the definition of overdensity significance, and the initial
follow-up spectroscopic observations are presented in \citet{toshikawa12}.
We give here a brief outline of our previous study.
We selected $i'$-dropout galaxies with $i'-z'>1.5$ in the wide field with $34\times27\,\mathrm{arcmin^2}$
field-of-view (FoV) of the SDF, and found a $6\sigma$ significance overdense region based on the surface number
density of the $i'$-dropout galaxies.
The region measures $\sim6\times6\,\mathrm{arcmin^2}$ over which the overdensity is more than $3\sigma$ significant
and includes $\sim30$ $i'$-dropout galaxies down to the $z'$-band limiting magnitude of $z'=27.09\,\mathrm{mag}$
($3\sigma$,$2\arcsec$ aperture).
Next, we targeted 31 $i'$-dropout galaxies for spectroscopy out of $\sim40$ $i'$-dropout galaxies in/around
the protocluster region.
Additionally, four galaxies were spectroscopically observed as secondary targets which do not meet our
photometric criteria perfectly but have $i-z>1.0$.
Totally, 35 galaxies were observed by follow-up spectroscopy, and we measured spectroscopic redshifts for
15 galaxies through their Ly$\alpha$ emissions.
Among these, eight galaxies are located within a narrow redshift range of $z=6.01\pm0.025$, and seven galaxies are
outside of this redshift range.
No spectroscopic redshifts were obtained for 20 galaxies.
This is presumably because these galaxies have faint/no Ly$\alpha$ emissions.
The contamination rate of dwarf stars and lower redshift galaxies was estimated to be about 6\%, and no apparent
contamination was found in our spectroscopically observed objects.

In this study, we performed further follow-up spectroscopy to completely observe $i'$-dropout galaxies in the
protocluster region; this deeper observation was conducted to detect faint Ly$\alpha$ emission lines that might
have been missed in the previous spectroscopic observations.
We also targeted the surrounding area of the protocluster region to seek for large-scale structures linked with
the protocluster.

\begin{figure}
\epsscale{1.1}
\plotone{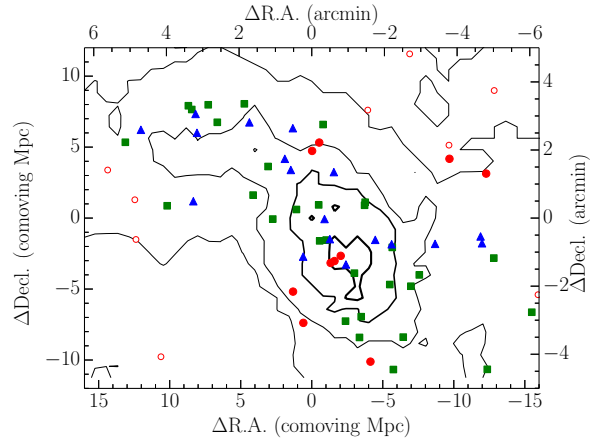}
\caption{Sky distribution of the $i'$-dropout galaxies and number density contours.
    Spectroscopically observed galaxies are marked by filled symbols (red circles: protocluster galaxies, blue
    triangles: field galaxies, green squares: Ly$\alpha$ undetected objects), and spectroscopically unobserved
    galaxies are shown by open circles.
    The origin (0,0) is $(\mathrm{R.A.}, \mathrm{Decl.})=(13:24:29.0,+27:17:19.1)$, which is defined as the
    center of the figure.
    The lines show the number density contours of $i'$-dropout galaxies from $6\sigma$ to $0\sigma$ with a step
    of $2\sigma$.
    The local number density was estimated by counting $i'$-dropout galaxies within a fixed aperture of $2\farcm1$
    radius ($5\,\mathrm{Mpc}$).
    $\sigma$ is the standard deviation of the local number densities.
    It can be clearly seen that all $i'$-dropout galaxies in/around the overdense region were spectroscopically
    observed.
    \label{sky}}
\end{figure}

\begin{figure*}
\epsscale{1.0}
\plotone{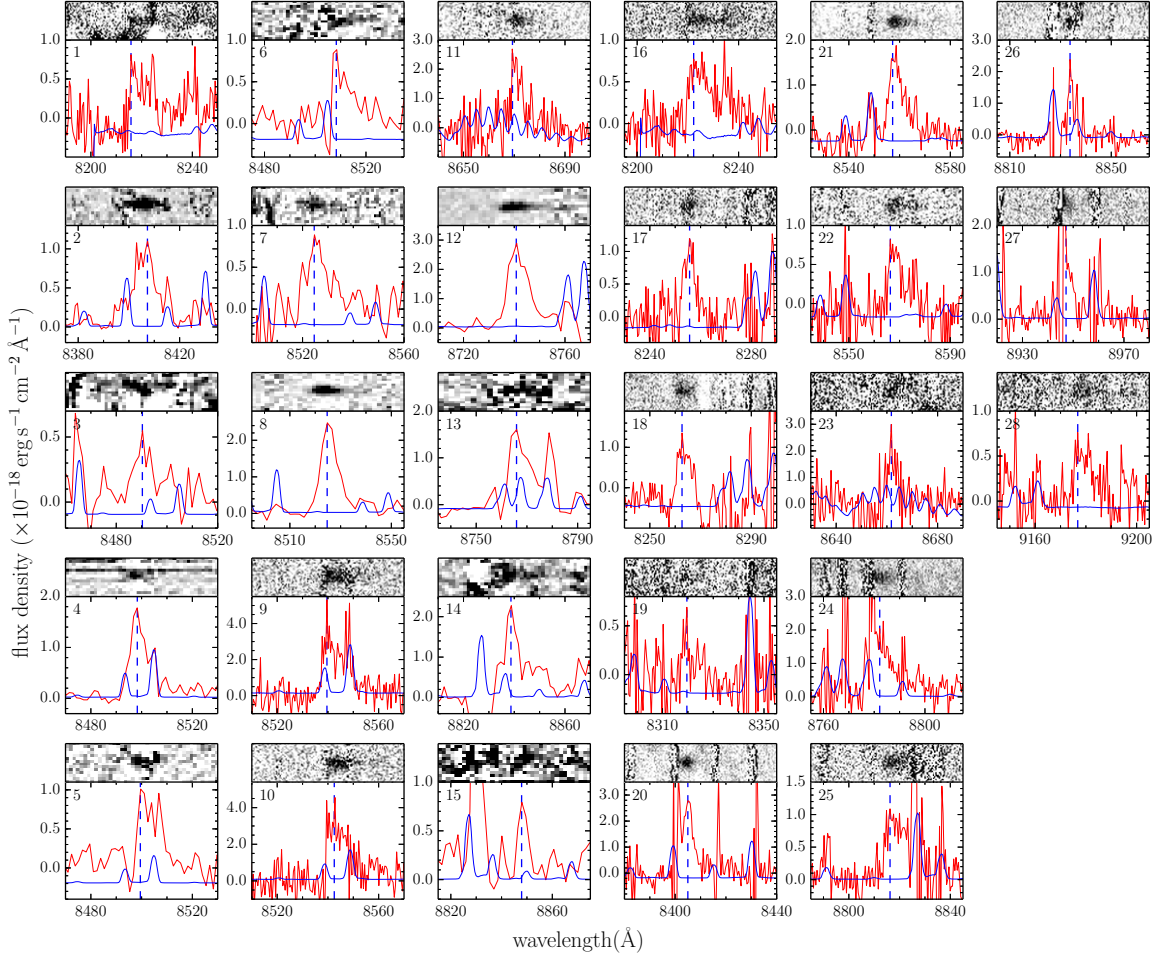}
\caption{Spectra of 28 galaxies having a Ly$\alpha$ emission line.
    The vertical dashed lines indicate the peak of the Ly$\alpha$ emission line.
    The blue solid lines represent the sky lines.
    The object IDs are indicated at the upper left corner (Column 1 of Table \ref{spec_cat}).
    \label{spec_fig}}
\end{figure*}

\subsection{Complete Follow-up Spectroscopy of the Protocluster} \label{spec}
We used the DEIMOS on the Keck II telescope in Multi-Object Spectroscopy (MOS) mode \citep{faber03}.
The DEIMOS observation was conducted using a $830\,\mathrm{line\,mm^{-1}}$ grating and the OG550 order blocking
filter, providing a wavelength coverage of $7000\,\mathrm{\AA}-10000\,\mathrm{\AA}$ with a pixel resolution of
$0.47\,\mathrm{\AA\,pix^{-1}}$.
The slit width was set to $1\farcs0$, giving a spectral resolution of $4.3\,\mathrm{\AA}$ ($R\sim2000$).
We targeted 34 $i'$-dropout galaxies in/around the protocluster; 14 of these were already observed by our previous
follow-up spectroscopy but no signal was detected.
In addition, we allocated slits to eleven secondary targets having $i'-z'>1.0$.
The observation was carried out on April 8, 2013, and the sky condition was good with a seeing of $\sim0\farcs6$.
We obtained 15 exposures for a total integration time of 7.5 hours.
Long slit exposures of a standard star, BD$+$332642, were used for the flux calibration.
The data were reduced using the pipeline spec2d\footnote{The data reduction pipeline was developed at the
University of California, Berkeley, with support from National Science Foundation grant AST 00-71048.}.
Furthermore, we obtained spectra of two $i'$-dropout galaxies through other MOS
observations with Keck/DEIMOS on different projects \citep{kashikawa11}.

Combining our previous follow-up spectroscopic observations, we observed 53 $i'$-dropout galaxies in/around the
protocluster region.
The DEIMOS pointing of this observation was set to cover both the protocluster region and a $2\sigma$ overdense
region extending from the main protocluster region.
The sky distribution of $i'$-dropout galaxies and spectroscopically observed galaxies are shown in Figure \ref{sky}.
All of the $i'$-dropout galaxies in the $2\sigma$ significant overdense region were completely observed with
spectroscopy.

\begin{turnpage}
\begin{deluxetable*}{ccccccccccc}
\tabletypesize{\scriptsize}
\tablecaption{Observed Properties of the 28 Spectroscopic Confirmed Galaxies \label{spec_cat}}
\tablewidth{0pt}
\tablehead{
    \colhead{ID\tablenotemark{a}} & \colhead{R.A.} & \colhead{Decl.} & \colhead{$z'$} & \colhead{$i'-z'$} &
        \colhead{$z$\tablenotemark{b}} & \colhead{$M_\mathrm{UV}$\tablenotemark{c}} &
        \colhead{$f_\mathrm{Ly\alpha}$\tablenotemark{d}} & \colhead{$L_\mathrm{Ly\alpha}$} &
        \colhead{$EW_\mathrm{rest}$\tablenotemark{e}} & \colhead{$S_w$} \\
    \colhead{} & \colhead{(J2000)} & \colhead{(J2000)} & \colhead{(mag)} & \colhead{(mag)} & \colhead{} &
        \colhead{(mag)} & \colhead{($10^{-18}\,\mathrm{erg\,s^{-1}\,cm^{-2}}$)} &
        \colhead{($10^{42}\,\mathrm{erg\,s^{-1}}$)} & \colhead{(\AA)} & \colhead{(\AA)}
}
\startdata
1 & 13:24:31.8 & +27:18:44.2 & $25.91\pm0.04$ & 1.48 & $5.758^{+0.001}_{-0.001}$ & $-20.74\pm0.04$ & $4.39\pm0.86$ & $1.59\pm0.31$ & $7.7\pm1.4$ & $5.61\pm1.36$ \\ 
2 & 13:24:18.4 & +27:16:32.6 & $25.69\pm0.03$ & 1.70 & $5.916^{+0.002}_{-0.002}$ & $-20.97\pm0.03$ & $12.49\pm0.36$ & $4.82\pm0.14$ & $19.0\pm0.8$ & $5.66\pm1.14$ \\ 
3 & 13:24:25.2 & +27:16:12.2 & $27.02\pm0.09$ & 1.89 & $5.984^{+0.006}_{-0.002}$ & $-19.60\pm0.10$ & $4.95\pm0.51$ & $1.96\pm0.20$ & $29.6\pm4.1$ & $3.41\pm4.03$ \\ 
4 & 13:24:30.2 & +27:14:13.5 & $26.81\pm0.08$ & 1.96 & $5.991^{+0.002}_{-0.002}$ & $-19.51\pm0.11$ & $15.42\pm0.63$ & $6.12\pm0.25$ & $107.5\pm12.2$ & $5.26\pm1.21$ \\ 
5 & 13:24:26.0 & +27:16:03.0 & $26.50\pm0.06$ & 1.71 & $5.992^{+0.002}_{-0.002}$ & $-20.11\pm0.10$ & $8.16\pm0.58$ & $3.24\pm0.23$ & $31.4\pm3.8$ & $10.01\pm0.93$ \\ 
6 & 13:24:21.3 & +27:13:04.8 & $25.91\pm0.04$ & 2.65 & $5.999^{+0.002}_{-0.006}$ & $-20.79\pm0.04$ & $7.59\pm0.68$ & $3.02\pm0.27$ & $15.3\pm1.5$ & $12.28\pm1.84$ \\ 
7 & 13:24:29.0 & +27:19:18.0 & $26.50\pm0.06$ & 1.60 & $6.012^{+0.002}_{-0.002}$ & $-20.11\pm0.07$ & $7.87\pm1.50$ & $3.15\pm0.60$ & $30.7\pm6.2$ & $5.06\pm1.84$ \\ 
8 & 13:24:28.1 & +27:19:32.8 & $26.10\pm0.04$ & 1.54 & $6.012^{+0.002}_{-0.002}$ & $-20.32\pm0.06$ & $21.34\pm0.57$ & $8.54\pm0.23$ & $70.8\pm4.5$ & $4.07\pm0.73$ \\ 
9 & 13:24:26.5 & +27:15:59.7 & $25.47\pm0.03$ & 1.91 & $6.025^{+0.003}_{-0.004}$ & $-21.03\pm0.05$ & $29.94\pm2.63$ & $12.04\pm1.06$ & $50.9\pm5.1$ & $8.00\pm0.21$ \\ 
10 & 13:24:31.5 & +27:15:08.8 & $25.91\pm0.06$ & 1.95 & $6.027^{+0.002}_{-0.002}$ & $-20.43\pm0.10$ & $27.36\pm2.95$ & $11.02\pm1.19$ & $95.7\pm14.2$ & $8.12\pm0.23$ \\ 
11 & 13:24:26.1 & +27:18:40.5 & $26.59\pm0.06$ & 2.32 & $6.131^{+0.003}_{-0.003}$ & $-19.87\pm0.11$ & $12.38\pm1.19$ & $5.19\pm0.50$ & $67.0\pm9.5$ & $4.24\pm0.35$ \\ 
12 & 13:24:31.6 & +27:19:58.2 & $26.47\pm0.06$ & 2.07 & $6.190^{+0.002}_{-0.002}$ & $-19.36\pm0.19$ & $24.48\pm0.69$ & $10.49\pm0.29$ & $213.1\pm40.3$ & $4.69\pm0.88$ \\ 
13 & 13:24:44.3 & +27:19:50.0 & $26.43\pm0.06$ & 2.49 & $6.211^{+0.004}_{-0.007}$ & $-20.23\pm0.09$ & $15.01\pm0.74$ & $6.48\pm0.32$ & $58.7\pm6.1$ & $7.24\pm1.24$ \\ 
14 & 13:24:20.6 & +27:16:40.5 & $27.01\pm0.09$ & 1.90 & $6.271^{+0.002}_{-0.005}$ & $-19.35\pm0.22$ & $14.28\pm0.94$ & $6.30\pm0.42$ & $136.4\pm32.1$ & $6.25\pm0.62$ \\ 
15 & 13:24:32.6 & +27:19:04.0 & $25.54\pm0.03$ & 1.47 & $6.278^{+0.005}_{-0.002}$ & $-21.75\pm0.03$ & $4.00\pm0.50$ & $1.77\pm0.22$ & $2.2\pm0.3$ & $2.35\pm3.20$ \\ 
16 & 13:24:44.8 & +27:17:48.8 & $26.10\pm0.04$ & 1.76 & $5.764^{+0.001}_{-0.001}$ & $-20.55\pm0.04$ & $7.08\pm0.72$ & $2.57\pm0.26$ & $16.0\pm1.8$ & $6.82\pm0.71$ \\ 
17 & 13:24:44.5 & +27:20:23.4 & $26.70\pm0.07$ & 1.09 & $5.791^{+0.001}_{-0.001}$ & $-19.95\pm0.07$ & $6.18\pm0.71$ & $2.27\pm0.26$ & $26.6\pm3.3$ & $2.80\pm1.39$ \\ 
18 & 13:24:27.4 & +27:17:17.4 & $26.62\pm0.06$ & 1.75 & $5.797^{+0.001}_{-0.001}$ & $-20.03\pm0.06$ & $5.67\pm0.70$ & $2.09\pm0.26$ & $23.3\pm2.9$ & $2.15\pm0.42$ \\ 
19 & 13:24:26.6 & +27:16:41.7 & $26.84\pm0.08$ & 1.74 & $5.844^{+0.001}_{-0.001}$ & $-19.84\pm0.08$ & $1.26\pm0.33$ & $0.47\pm0.12$ & $6.3\pm1.5$ & $-0.66\pm0.67$ \\ 
20 & 13:24:12.7 & +27:16:33.3 & $25.99\pm0.04$ & 1.39 & $5.914^{+0.001}_{-0.001}$ & $-20.67\pm0.04$ & $13.17\pm0.74$ & $5.08\pm0.29$ & $30.2\pm2.2$ & $2.51\pm0.27$ \\ 
21 & 13:24:10.8 & +27:19:04.0 & $26.61\pm0.06$ & 2.30 & $6.039^{+0.001}_{-0.001}$ & $-19.88\pm0.09$ & $10.95\pm0.99$ & $4.43\pm0.40$ & $54.6\pm8.3$ & $6.04\pm0.57$ \\ 
22 & 13:24:05.9 & +27:18:37.7 & $26.87\pm0.08$ & 2.04 & $6.047^{+0.001}_{-0.001}$ & $-19.84\pm0.09$ & $3.97\pm0.72$ & $1.61\pm0.29$ & $20.9\pm4.5$ & $4.53\pm0.82$ \\ 
23 & 13:24:30.2 & +27:16:10.8 & $26.15\pm0.05$ & 2.21 & $6.125^{+0.001}_{-0.001}$ & $-20.61\pm0.06$ & $9.37\pm0.60$ & $3.92\pm0.25$ & $28.4\pm2.4$ & $0.93\pm0.19$ \\ 
24 & 13:24:37.3 & +27:20:08.5 & $26.00\pm0.04$ & 2.88 & $6.224^{+0.001}_{-0.001}$ & $-20.93\pm0.06$ & $13.46\pm1.37$ & $5.85\pm0.59$ & $26.9\pm3.9$ & $30.20\pm6.33$ \\ 
25 & 13:24:51.7 & +27:19:55.3 & $26.84\pm0.08$ & 2.07 & $6.252^{+0.001}_{-0.001}$ & $-19.99\pm0.12$ & $8.90\pm0.57$ & $3.90\pm0.25$ & $43.1\pm6.6$ & $0.87\pm0.90$ \\ 
26 & 13:24:06.6 & +27:16:46.4 & $26.81\pm0.08$ & 1.79 & $6.267^{+0.001}_{-0.001}$ & $-20.28\pm0.09$ & $4.99\pm0.40$ & $2.20\pm0.18$ & $19.3\pm2.5$ & $2.00\pm0.11$ \\ 
27 & 13:24:24.5 & +27:15:56.9 & $26.69\pm0.07$ & 2.22 & $6.359^{+0.001}_{-0.001}$ & $-20.25\pm0.12$ & $14.07\pm0.71$ & $6.41\pm0.32$ & $58.1\pm8.4$ & $8.28\pm2.03$ \\ 
28 & 13:24:06.5 & +27:16:34.3 & $26.97\pm0.09$ & 1.94 & $6.549^{+0.001}_{-0.001}$ & $-20.92\pm0.10$ & $4.39\pm0.64$ & $2.14\pm0.31$ & $9.3\pm2.0$ & $3.86\pm1.35$ \\ 
\enddata
\tablenotetext{a}{The galaxies with ID=1-15 were identified by previous observations, and have the same
    ID as in Table 3 of \citet{toshikawa12}.
    The others (ID=16-28) were newly identified in this observation.}
\tablenotetext{b}{The redshifts were derived by the peak wavelength of the Ly$\alpha$ emission line, assuming
    the rest wavelength of Ly$\alpha$ to be 1215.6{\AA}.
    These measurements could be overestimated in the case of significant damping wings by IGM.
    When emission lines are located near strong sky lines, line profiles are affected by them, and the position of
    the peak could be shifted.
    These effects of sky lines and the wavelength resolution are taken into account estimating the error.}
\tablenotetext{c}{The absolute magnitudes of the UV continuum (at 1300{\AA} in the rest-frame) were
    derived from $z'$-band magnitudes using equation (1) of \citet{kashikawa11}, assuming that the UV
    continuum slopes ($f_\lambda \propto \lambda^\beta$) are $\beta = -2$, and the continua blueward of
    Ly$\alpha$ are entirely absorbed by intergalactic medium.
    No aperture correction was applied.
    The error was estimated using the following Monte Carlo simulation: 1) Gaussian random errors were
    assigned to the measured Ly$\alpha$ flux and $z'$-band magnitude and $M_\mathrm{UV}$ was re-calculated;
    2) the process is repeated 100,000 times; 3) the error of $M_\mathrm{UV}$ was determined from the rms
    fluctuation.
    The systematic error depending on the assumption of $\beta$ was not taken into account, since the scatter of
    $i'$-dropout galaxies' $\beta$ is not well constrained \citep[e.g.,][]{bouwens12,wilkins11}.
    Even if $\beta$ was varied from -3.0 to -1.0, the variation of the UV continuum flux is a few percent.}
\tablenotetext{d}{The observed line flux corresponds to the total amount of the flux within the line profile.
    The error was estimated from the noise level at wavelengths blueward of Ly$\alpha$.}
\tablenotetext{e}{The rest-frame equivalent widths of Ly$\alpha$ emission were derived from the combination
    of $M_\mathrm{UV}$ and $L_\mathrm{Ly\alpha}$.
    The error was estimated from $M_\mathrm{UV}$, $L_\mathrm{Ly\alpha}$ and their errors.
    The systematic error of $\beta$ is not taken into account.}
\end{deluxetable*}
\end{turnpage}

\section{RESULTS} \label{result}
From the observation as shown in above section, 14 emission lines are identified by careful examination of
two- and one-dimensional spectra.
All emission lines in our wavelength coverage are single emission lines; thus, there is almost no chance that
H$\beta$ or [\ion{O}{3}] emission lines contaminate our spectra because the wavelength coverage of our observation
is wide enough to detect all these lines simultaneously.
However, only [\ion{O}{3}]$\lambda5007$ emission, which is typically the strongest emission among them, may be
detected if the other emissions are too faint to detect.
We investigated the possibility that H$\alpha$, [\ion{O}{2}], and [\ion{O}{3}]$\lambda5007$ emission lines might
have contaminated our sample, based on both imaging and spectroscopic data.
\citet{haines08} demonstrated that $\sim30\%$ of red-sequence galaxies in the field have ongoing star-formation
activity with $\mathrm{EW(H\alpha)}>2\,\mathrm{\AA}$, but they also found that these galaxies disappear at an
absolute $r$-band magnitude of $M_r\gtrsim-18$.
Our samples are much fainter ($>3\,\mathrm{mag}$) than this magnitude, if they were at $z\sim0.3$ based on the
photometry of our samples.
Regarding [\ion{O}{2}] doublet emission lines, it is possible to distinguish between Ly$\alpha$ and [\ion{O}{2}]
emission lines based on the line profile.
The spectral resolution of our spectroscopic observation is high enough to resolve [\ion{O}{2}] emission lines
as doublets ($\Delta\lambda=6.3\,\mathrm{\AA}$ at $z\sim1.3$), although it would be practically difficult to
resolve these in most cases due to low signal-to-noise ratio (S/N).
In this case, the [\ion{O}{2}] emission line should be skewed to blueward, while Ly$\alpha$ emission lines from
high redshift galaxies should be skewed to redward.
We also calculated weighted skewness, $S_w$, which is a good indicator of the line asymmetry \citep{kashikawa06}.
The asymmetric emission lines with $S_w>3$ are evidence of Ly$\alpha$ emission from high redshift
galaxies.
Although some emission lines of this study have $S_w<3$, strong sky lines and low S/N data prevent the accurate
determination of skewness in these cases.
The red optical color of $i'-z'>1.5$ may indicate passive or dustier galaxies if they are at $z\sim1.5$ or 0.7,
while [\ion{O}{2}] and [\ion{O}{3}]$\lambda5007$ emissions contradict with the prominent emission lines as the
sign of high star-formation activity.
Furthermore, according to \citet{ly07,ly12}, [\ion{O}{2}] emitters at $z\sim1.5$ and [\ion{O}{3}] emitters at
$z\sim0.7$ typically have $i'-z'\sim0.2$, and almost all have $i'-z'<1.0$, much bluer than the selection criterion
of our sample ($i'-z'>1.5$).
Therefore, it is unlikely that H$\alpha$, [\ion{O}{2}], and [\ion{O}{3}]$\lambda$5007 emission lines contaminate
our sample, and Ly$\alpha$ is the most plausible interpretation for these single emission lines.
We can conclude that  all single emission lines detected from our $i'$-dropout sample as Ly$\alpha$ emission lines
at $z\sim6$.

Although the detection limit of an emission line depends on the wavelength and line width, the $1\sigma$ detection
limit for emission lines is typically $6.3\times10^{-19}\,\mathrm{erg\,s^{-1}\,cm^{-2}}$ in this observation,
which was derived from the sky fluctuations in a typical line width ($\sim6.3\,\mathrm{\AA}$).
The limiting observed-frame equivalent width corresponding to the line detection limit is estimated to be 
$26.4\,\mathrm{\AA}$ ($5.3\,\mathrm{\AA}$) at $m_{z'}=27.0\,\mathrm{mag}$ ($25.0\,\mathrm{mag}$).

\begin{figure}
\epsscale{1.1}
\plotone{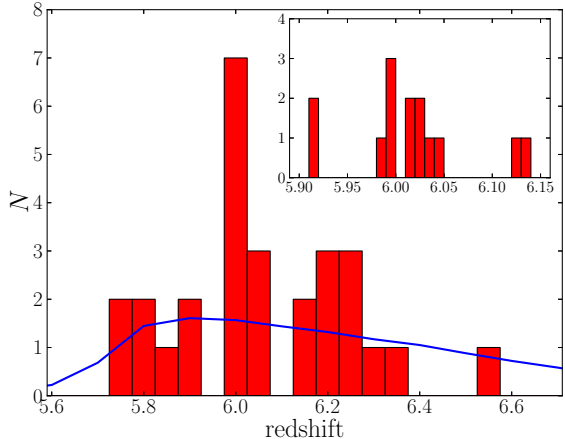}
\caption{Redshift distribution of the 28 spectroscopically confirmed galaxies.
    The bin size is $\Delta z = 0.05$.
    The solid (blue) line shows the selection function of our $i'$-dropout selection assuming a uniform 
    distribution normalized to the total number of confirmed emitters.
    The inset is a close-up of the protocluster redshift range, with a bin size of $\Delta z =0.01$.
    \label{redshift}}
\end{figure}

One of the targets in this study was already detected as ID1 of \citet{toshikawa12}, whose S/N was worse
in the previous study; thus, the spectroscopic properties of ID1 are replaced with those derived in this study.
Therefore, we were able to newly confirm 13 $z\sim6$ galaxies; thus, in total 28 $z\sim6$ galaxies, including the
15 in \citet{toshikawa12}, are identified in/around the overdense region.
We estimated the observed properties of the spectroscopically confirmed galaxies, such as UV absolute magnitude
($M_\mathrm{UV}$), Ly$\alpha$ luminosity ($L_\mathrm{Ly\alpha}$), and rest-frame Ly$\alpha$ equivalent width
($EW_\mathrm{rest}$).
Although faint continuum flux could not be detected in the observed spectra, $M_\mathrm{UV}$ were estimated from
the $z'$-band magnitudes by subtracting the spectroscopically measured Ly$\alpha$ flux and assuming flat UV
continuum spectra ($f_\nu=\mathrm{constant}$).
The spectra and observed properties of all these galaxies are provided in Figure \ref{spec_fig} and Table
\ref{spec_cat}.
The redshift distribution is shown in Figure \ref{redshift}.
It is clear that ten galaxies are clustered in a narrow redshift range between $z=5.984$ and $z=6.047$
($\Delta z\lesssim0.06$), corresponding to the radial distance of $26.1\,\mathrm{Mpc}$ in comoving units.
The central redshift of the protocluster is estimated to be $z=6.01$ using biweight \citep{beers90} of ten
galaxies.
This concentration is about 4.5 times higher than the number expected from a homogeneous distribution in redshift
space.

\section{COMPARISON WITH MODEL PREDICTIONS} \label{model}
We found a protocluster with $6\sigma$ overdensity at $z=6.01$, but how large is the dark matter halo that it will
evolve into at $z=0$?
To answer this question, it is necessary to compare our observations with theoretical predictions about the
descendants of high redshift overdensities by tracing hierarchical merging histories.
\citet{overzier09} and \citet{chiang13} investigated the relation between galaxy overdensity at high redshift
and dark matter halo mass at $z=0$ by using a combination of $N$-body dark matter simulations and semi-analytic
galaxy formation models.
They systematically studied cluster development from $z\sim6$ to $z=0$ and found clear correlations between
overdensity at high redshift and halo mass at $z=0$, depending on e.g., the sample selection, search volume, and
redshift accuracy of the tracer galaxies, as well as the mass of the clusters.
Here, we perform a new simulation specifically designed to match the observational details of our $i'$-dropout
survey as closely as possible.
This design is important because our target selection is characterized by a rather broad photometric selection at
$z=6\pm0.5$ compared with the cases described by \citet{chiang13}.
We will connect directly the observed quantity, the significance of the overdensity of the surface number density,
to the dark matter halo mass at $z=0$.

\begin{figure}
\epsscale{1.1}
\plotone{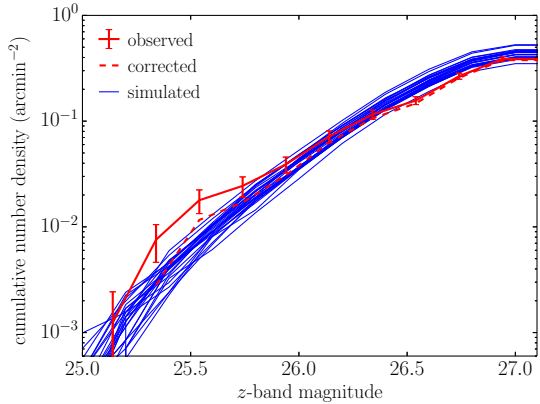}
\caption{Number counts of the $i'$-dropout galaxies.
    The red and blue solid lines indicate the observed galaxies in the SDF and those from in the 24 light-cone
    models, respectively.
    The number counts of the SDF were corrected for incompleteness.
    The red dashed line shows the observed number counts corrected for contamination by dwarf stars by using the
    star count model \citep{nakajima00}.
    \label{num_count}}
\end{figure}

We used the light-cone model constructed by \citet{henriques12}.
A brief outline of the construction of light-cone models is presented below.
First, the assembly history of the dark matter halos was traced using an $N$-body simulation \citep{springel05},
in which the length of the simulation box was $500\,h^{-1}\,\mathrm{Mpc}$ and the particle mass was
$8.6\times10^8\,h^{-1}\,\mathrm{M_\sun}$.
The distributions of dark matter halos were stored at discrete epochs.
Next, the processes of baryonic physics were added to dark matter halos at each epoch using a semi-analytic galaxy
formation model \citep{guo11}.
Based on the intrinsic parameters of galaxies predicted by the semi-analytic model, such as stellar mass, star
formation history, metallicity, and dust content, the photometric properties of simulated galaxies were estimated 
from the stellar population synthesis models developed by \citet{BC03}.
Then, these simulated galaxies in boxes at different epochs were projected along the line-of-sight, and
intergalactic medium (IGM) absorption was applied in order to mimic a pencil-beam survey using the \citet{madau95}
IGM light-cone set from \citet{overzier13}.
Finally, 24 light-cone models with $1.4\times1.4\,\mathrm{deg^2}$ FoV were extracted using these procedures.

\begin{figure}
\epsscale{1.1}
\plotone{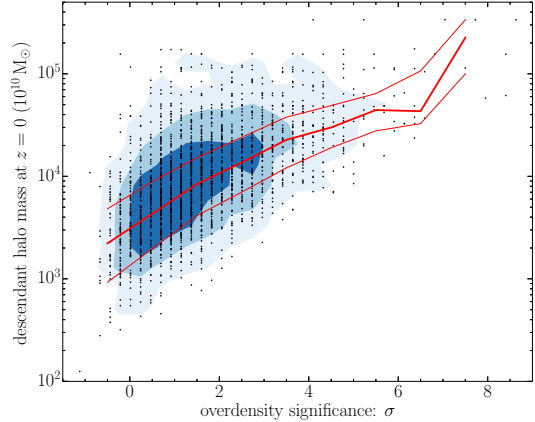}
\caption{Relation between surface overdensity at $z\sim6$ and descendant halo mass at $z=0$.
    The points represent descendant halo masses in each overdense region.
    The thick and thin red lines are the median, upper, and lower quartiles.
    The background contours show the 50, 75, and $95\,\%$ region from dark to light.
    \label{z0mass}}
\end{figure}

The same color selection applied to the observations \citep{toshikawa12} was also applied to the simulated catalog
of each light-cone model.
In this analysis, we only used the galaxies whose dark matter halos are composed of more than 100 dark matter
particles ($\sim1\times10^{11}\,\mathrm{M_\sun}$ halo mass), because galaxies' SEDs in small halos are not stable
between redshift snapshots.
A color correction was also made to transform the light-cone's SDSS filter system to Subaru/SprimeCam's one.
We detected $\sim3000$ $i'$-dropout galaxies per $2\,\mathrm{deg^2}$ FoV down to the same limiting magnitude of
$z'$-band, as in the SDF.
The number counts of the simulated $i'$-dropouts were found to be nearly consistent with those observed in the
SDF (Figure \ref{num_count}).
Although there is some deviation at the bright end, this is probably caused by the contamination of dwarf stars,
which are not included in the model.
As in Figure \ref{num_count}, the observed number counts well agree with the simulated ones by correcting for the
contamination of dwarf stars, which is estimated by the star count model developed by \citet{nakajima00}.
We also compared the simulated number counts with those of $z\sim6$ galaxies in \citet{bouwens14}, applying the
selection window for $z\sim6$ galaxies presented in \citet{bouwens14} to light-cone models.
Both number counts are almost consistent with each other.
Next, we investigated the sky distribution and estimated the local surface number density in the light-cone
catalogs in the same way as in the SDF.
For each overdense region, the dominant structure was identified as follows.
First, we selected the strongest spike in the redshift distribution for that region.
Next, we determined the halo ID of the most massive halo in that redshift spike.
Finally, the descendant halos at $z=0$ for each overdense region were identified by tracing the halo merger tree of
those $z\sim6$ halos.
Figure \ref{z0mass} shows the relation between the significance of the overdensities at $z\sim6$ and the halo mass
at $z=0$.
Although there is large scatter, these two quantities are correlated quite closely.
The Spearman's rank correlation coefficient is 0.55, indicating that the probability of no correlation is
$<10^{-5}$.
We find that 95\% (100\%) of $>4\sigma$ ($>6\sigma$) overdense regions are expected to include protoclusters.
This result suggests that we can detect a real protocluster with high confidence by measuring the overdensity
significance if it is more than $4\sigma$ away from the observed surface number density, and also that we can
infer its descendant halo mass at $z=0$ based on Figure \ref{z0mass}.
The protocluster that we found with $6\sigma$ significance at $z\sim6$ is predicted to evolve into a dark matter
halo with a mass of $5\times10^{14}\,\mathrm{M_\sun}$ on average and at least $1\times10^{14}\,\mathrm{M_\sun}$
at $z=0$, suggesting that this is a progenitor of a local galaxy cluster.
We also estimated the number density of such overdense regions at $z\sim6$ based on the large survey volume covered
by the simulations.
The average number density of regions with $>6\sigma$ significance is $0.42\,\mathrm{deg^{-2}}$ (0.83 per
light-cone), though the scatter from field to field is as large as $0.43\,\mathrm{deg^{-2}}$.
Considering that our finding was made in a field measuring only $0.24\,\mathrm{deg^2}$, this discovery must have
been quite serendipitous.

\begin{figure}
\epsscale{1.1}
\plotone{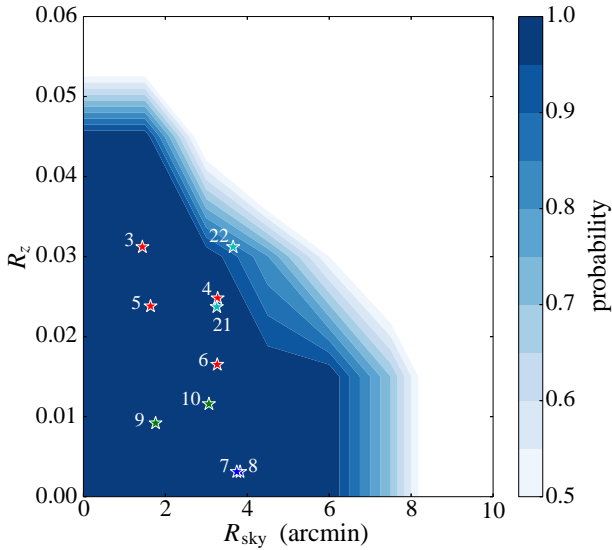}
\caption{Probability of merging into a single halo by $z=0$ as a function of distance from the center of the
    protoclusters at $z\sim6$.
    Probability was defined as the fraction of protocluster members among all galaxies at that distance.
    This probability map was derived from stacking all $>6\sigma$ regions after calculating the probability map
    of each overdense region.
    The horizontal and vertical axes indicate spatial and redshift directions, and color contours show the
    probability.
    The positions of our ten protocluster members (ID=3-10,21,22) are also plotted using the same color code as
    that used in Figure \ref{3D}.
    The figure shows that is is highly likely that the ten spectroscopically confirmed galaxies will
    all end up within the same cluster-sized halo at $z=0$.
    \label{p_member}}
\end{figure}

In addition, we estimated the probability for a galaxy located at a certain position relative to the protocluster
center to become a cluster member at $z=0$.
Based on the simulation, we can map the galaxy distribution of a protocluster at $z\sim6$ by tracing back the
merger tree of a cluster at $z=0$.
Although protoclusters have different structural morphologies, such as filamentary or sheet-like, we estimated
the probability by counting the numbers of both descendant cluster members and non-members as a function of the
distance to the protocluster center.
The center of each protocluster is defined as the peak of the three-dimensional density field, and a spatial map
of these probabilities is derived for each protocluster.
We extracted 19 $z\sim6$ protoclusters with $>6\sigma$ overdensity and made a median stack in order to derive a
spatial map of the probability that a galaxy will merge into a single halo by $z=0$ as a function of distance from
the center of the protocluster (Figure \ref{p_member}).
Additionally, we plotted the observed protocluster members within a narrow redshift range (ID=3-10,21,22) measured
by our spectroscopy, as shown in Figure \ref{p_member}.
The center of the protocluster is simply defined by the middle values between the maximum and the minimum of
R.A./Decl./redshift of ten galaxies, which is around $(\Delta\mathrm{R.A.},\Delta\mathrm{Decl.}) =
(-4,-2)\,\mathrm{Mpc}$ in Figure \ref{sky} and $z=6.02$.
We found that these ten (nine without ID=22) galaxies will become a single halo by $z=0$ with a probability of
$>80\,\%$ ($>95\,\%$).
Based on the model prediction, even two galaxies, which are located around $(\Delta\mathrm{R.A.},
\Delta\mathrm{Decl.})=(-11,3)\,\mathrm{Mpc}$ in Figure \ref{sky}, are expected to coalesce into a cluster even
though they are located in a region of only $<2\sigma$ significance.
This illustrates well that our ten protocluster members are indeed expected to become members of a cluster at $z=0$.
It should be noted that the estimate of the center of the observed protocluster involves some uncertainties,
because only Ly$\alpha$ emitting galaxies can be identified as protocluster members; detection completeness also
grows some uncertainties.
In this study, the number of protocluster members is only ten; therefore, we estimated the center of the observed
protocluster in various ways, using the average, median, or biweight \citep{beers90} of the ten members' positions,
and checked the differences between these methods.
As a result, the uncertainties in the center are $<1.5\,\mathrm{arcmin}$ ($3.6\,\mathrm{Mpc}$) in the
spatial and $<0.01$ ($4.2\,\mathrm{Mpc}$) in the redshift direction.
Although the most outer galaxy could have a probability of only $50\,\%$ or less in the worst case, most galaxies
have a high probability of $>80\,\%$.
Therefore, we conclude that these ten galaxies are likely protocluster members.

\section{DISCUSSION} \label{discuss}
Based on comparison with the model prediction in the previous section, we classified 28 spectroscopically
confirmed galaxies into two groups: ten protocluster members (ID=3-10,21,22) and 18 non-members
(ID=1,2,11-20,23-28).
The members are clustered in the redshift range between $z=5.984$ and $z=6.047$ ($\Delta z\lesssim0.06$),
which corresponds to a radial separation of $26.1\,\mathrm{Mpc}$.
The properties and distribution of the galaxies within the protocluster are discussed in this section.

\subsection{Galaxy Properties}
\begin{figure}
\epsscale{1.1}
\plotone{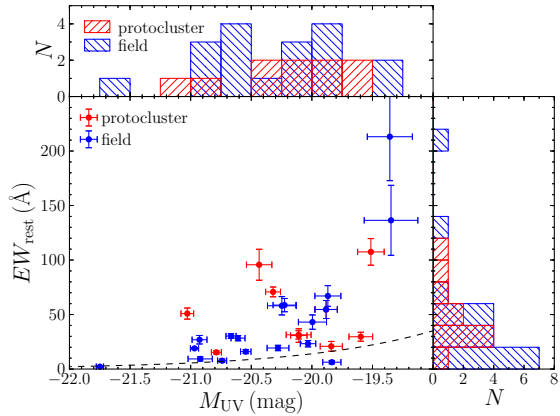}
\caption{$EW_\mathrm{rest}$ versus $M_\mathrm{UV}$ of spectroscopically confirmed galaxies.
    The histograms in the top and right panels show the $EW_\mathrm{rest}$ and $M_\mathrm{UV}$ distributions of
    the protocluster and field galaxies.
    Red and blue color represent the protocluster and field galaxies, respectively.
    The dashed line indicates the $5\sigma$ detection limit of the spectroscopic observation.
    \label{M-EW}}
\end{figure}

We compared several galaxy properties between members and non-members to investigate whether there was any
difference due to their environments at this early epoch.
The average and standard deviation of $L_\mathrm{Ly\alpha}$, $M_\mathrm{UV}$, and $EW_\mathrm{rest}$ in the
protocluster and field galaxies were estimated, and found that all of these properties of protocluster and field
galaxies are consistent with each other within $1\sigma$ scatter, as shown in Table \ref{ave_cat}.
Figure \ref{M-EW} shows that there are no significant differences between the $M_\mathrm{UV}$ and Ly$\alpha$
$EW_\mathrm{rest}$ distribution between protocluster and field (the p-value derived by the Kolmogorov-Smirnov (KS)
test is $>0.3$); however, a possible difference can be seen at the lowest $EW_\mathrm{rest}$ bin
($EW_\mathrm{rest}<20\,\mathrm{\AA}$).
Because both protocluster and field galaxies were observed in the same observing runs, it is unlikely that this
difference was caused by a difference in the completeness limit of the spectroscopic observations.
Taking into account the completeness limit, the sample in the lowest $EW_\mathrm{rest}$ bin mostly consists of
the brightest galaxies in $M_\mathrm{UV}$.
This may imply that the brightest galaxies in the overdense region are older and have more dust, suppressing the
Ly$\alpha$ emission compared with those in the field galaxies.
\citet{lee13} also reported that the median $EW_\mathrm{rest}$ of $z=3.78$ protocluster galaxies is higher than
that of field galaxies.
However, the Ly$\alpha$ emission, which is a resonantly-scattered line, can be affected by many physical parameters
such as SFR, dust amount, and geometry of dust and neutral gas \citep[e.g.,][]{shibuya14}.
It is therefore difficult to identify the reason of the possible difference at the lowest $EW_\mathrm{rest}$.
In this study of a $z=6.01$ protocluster, it should be noted that the difference can be only seen in the lowest
$EW_\mathrm{rest}$ bin, and there is no significant difference between the $EW_\mathrm{rest}$ distribution of the
protocluster and the field galaxies as a whole.

\begin{deluxetable}{cccc}
\tablecaption{Average of Observed Properties of Protocluster and Field Galaxies \label{ave_cat}}
\tablewidth{0pt}
\tablehead{
    \colhead{} & \colhead{$L_\mathrm{Ly\alpha}$} & \colhead{$M_\mathrm{UV}$} & \colhead{$EW_\mathrm{rest}$} \\
    \colhead{} & \colhead{($10^{42}\,\mathrm{erg\,s^{-1}}$)} & \colhead{(mag)} & \colhead{(\AA)}
}
\startdata
protocluster & $5.51\pm3.78$ & $-20.16\pm0.49$ & $50.74\pm31.66$ \\
field & $4.21\pm2.44$ & $-20.34\pm0.61$ & $45.53\pm52.60$ \\
\hline
p-value\tablenotemark{a} & 0.78 & 0.78 & 0.32
\enddata
\tablenotetext{a}{Using the KS test, the distribution of observed properties are compared between protocluster
    and field galaxies.}
\end{deluxetable}

We measured the Ly$\alpha$ fraction, which is the fraction of Ly$\alpha$ emitting galaxies among our dropout
galaxies.
This has been widely studied at $4<z<8$ \citep[e.g.,][]{stark10,schenker12,treu13}, and the fraction steadily
increases toward higher redshift, while it gradually decreases beyond $z=6$, possibly as a signature of
reionization.
However, those measurements were made using only field galaxies.
It is important to compare the Ly$\alpha$ fraction between field galaxies and galaxies in overdense regions in
order to ascertain whether it has environmental dependence or not.
Since it is impossible to distinguish between member and non-member galaxies in the spectroscopically undetected
galaxies, we measured the Ly$\alpha$ fraction in the projected overdense region over $8\times8\,\mathrm{arcmin^2}$,
including Ly$\alpha$ emitting galaxies, even if they were found in the field behind and in front of the
protocluster.
We assumed that the Ly$\alpha$ undetected galaxies are all at $z=6.0$, and calculated their $M_\mathrm{UV}$.
In this estimate, expected IGM absorption at $z=6.0$ in the $z'$-band was also corrected for though the correction
is as small as $0.08\,\mathrm{mag}$.
We here apply the same color selection criterion of $i'-z'>1.3$ as in \citet{stark11}.
Our spectroscopic completeness still remains high ($>95\%$) in the overdense region because $i'$-dropout galaxies
with $1.3<i'-z'<1.5$ were also observed as secondary targets.
We compare our result with the bright sample of \citet{stark11}, which has the same $M_\mathrm{UV}$ range as ours.
The fraction in the overdense region was found to be $0.0^{+6.4}_{-0.0}\%$ and $20.0\pm11.0\%$ for
$EW_\mathrm{rest}>50\,\mathrm{\AA}$ and $>25\,\mathrm{\AA}$, which are almost the same as in the field at $z\sim6$.

These results show that we do not find any significant differences in the observed properties between protocluster
and field galaxies.
This would indicate that this protocluster is still in the early phase of cluster formation, before any
environmental effect works on galaxy properties.
However, the observed properties of protocluster galaxies in this study are very limited.
According to other works for lower-redshift protoclusters, differences in galaxy properties between protocluster
and field galaxies begin to appear at $z\sim2-3$: protocluster galaxies are $\sim2-3$ times more massive than
field galaxies at the same redshift \citep{steidel05,kuiper10,hatch11}.
In this study, the UV and Ly$\alpha$ properties of the protocluster galaxies are almost the same as those of
field galaxies.
These properties are closely related to star-formation activity.
This could imply that galaxy mergers could scarcely happen even in the high-density region at $z\sim6$ because
galaxies basically contain large amounts of gas in their young phase.
Under such conditions, galaxy mergers can ignite bursty star formation.
Therefore, the stellar-mass difference as seen at $z=2-3$ could emerge in a later cluster formation phase rather
than in this early stage; overdense regions would result in a higher rate of galaxy mergers, which could prolong
star formation.
However, direct stellar mass measurements using deep infrared imaging of the protocluster at $z=6.01$ will be
required to investigate this issue further.

\begin{figure}
\epsscale{1.1}
\plotone{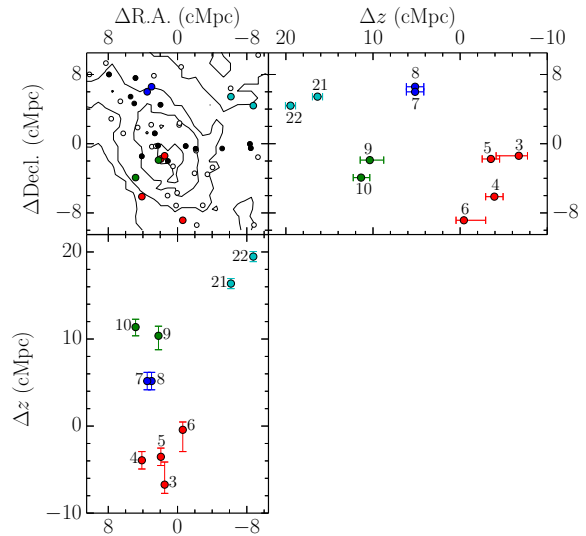}
\caption{Three-dimensional distribution of the protocluster galaxies.
    The points and lines are the $i'$-dropout galaxies and number density contours respectively, which are the same
    as those in Figure \ref{sky}.
    The filled points represent the 28 Ly$\alpha$ detected galaxies, and the color-coded points indicate the
    protocluster galaxies.
    Possible substructures are grouped by the same color
    Note that the origin (0,0) of this figure is defined as
    $(\mathrm{R.A.}, \mathrm{Decl.})=(13:24:22.4,+27:16:47.3)$, which is different from that of Figure \ref{sky}.
    \label{3D}}
\end{figure}

\subsection{Protocluster Structure}
\begin{figure}
\epsscale{1.16}
\plottwo{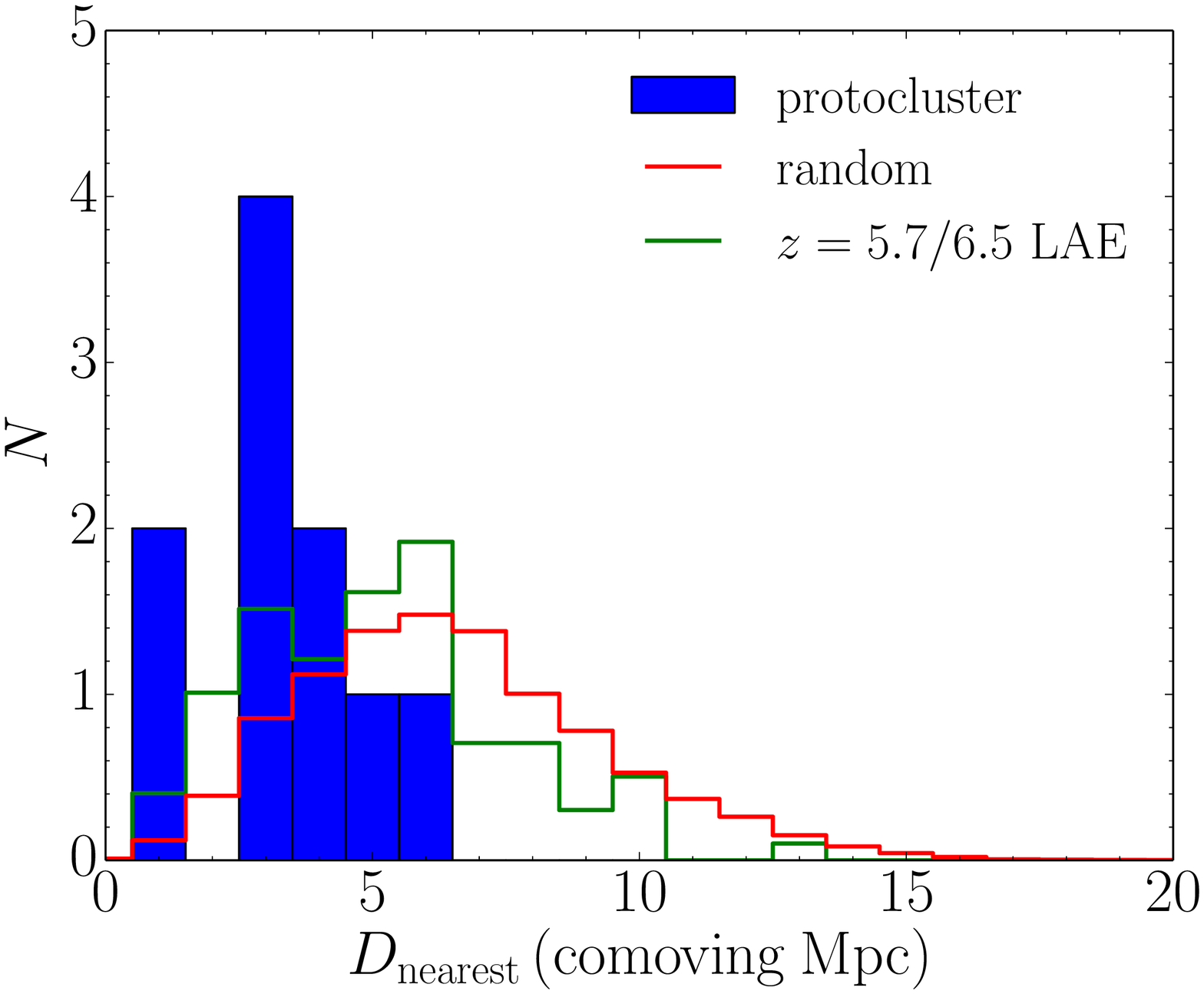}{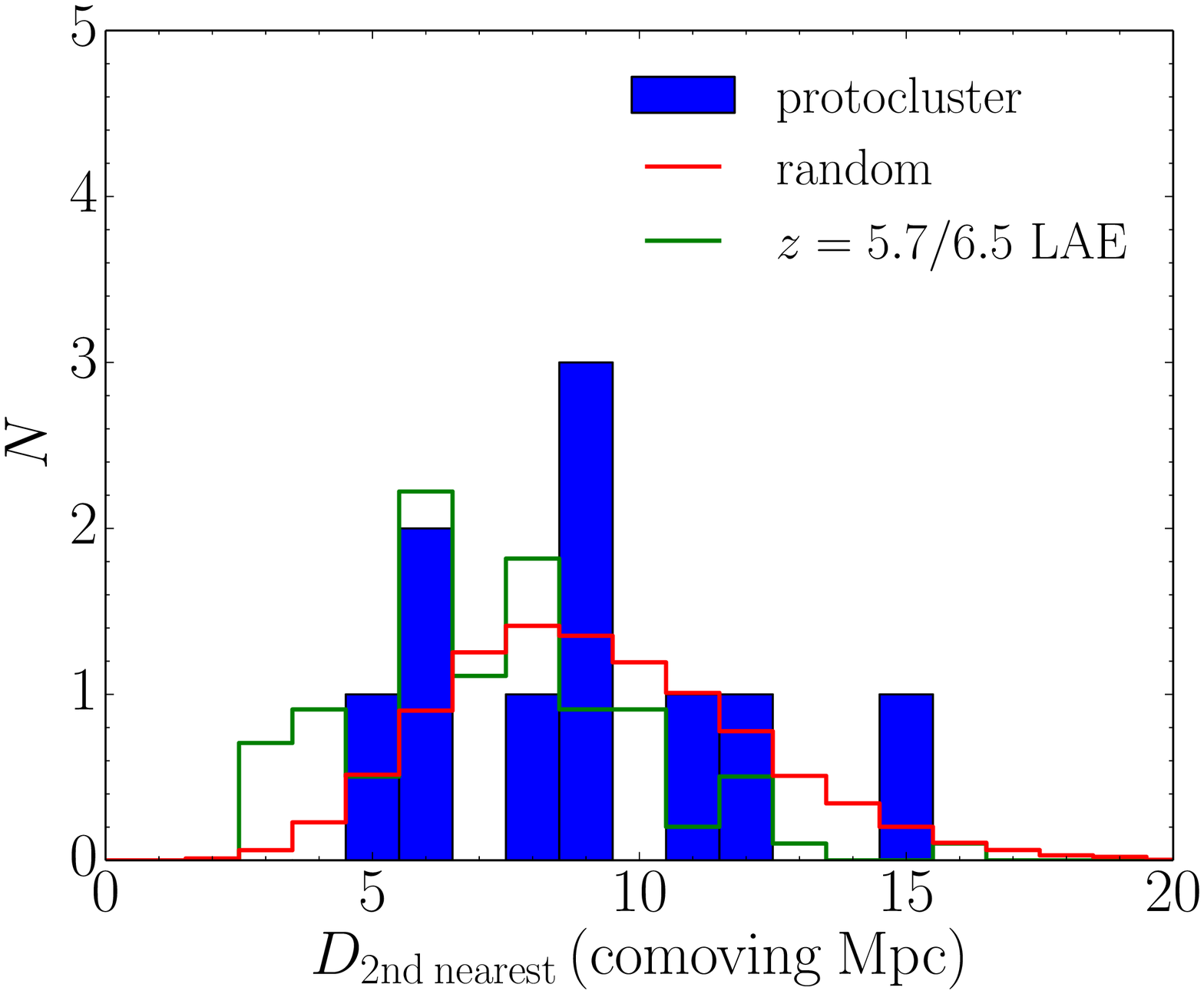}
\caption{Distribution of the separation from the first (left) and second (right) nearest galaxy in the
    protocluster (blue histogram).
    The red line shows an expected distribution assuming that ten galaxies are randomly distributed in the
    protocluster region of $17\times19\times29\,\mathrm{Mpc^3}$, which was defined by the smallest box,
    including ten protocluster galaxies.
    This random realization was repeated 1,000 times.
    The green line shows the distribution of $z=5.7$ and 6.5 LAEs in the SDF \citep{kashikawa11}.
    Separation of field LAEs is corrected for the difference in the average number density between the
    protocluster and field by multiplying $(n_\mathrm{field}/n_\mathrm{protocluster})^{1/3}\sim0.6$.
    \label{D1st}}
\end{figure}

As has been discussed in \citet{toshikawa12}, the protocluster appears to have some substructure: there are no
spectroscopically identified galaxies at the central region of the protocluster, and it is far from virial
equilibrium.
In this study, the number of spectroscopically identified member galaxies was increased to ten, but their velocity
dispersion was found to be still too high ($869\pm85\,\mathrm{km\,s^{-1}}$) to consider virial equilibrium.
Figure \ref{3D} presents an updated version of the 3D galaxy distribution in the protocluster after including new
spectroscopically identified galaxies; it reveals that the protocluster seems to consist of 4 subgroups of close
pairs.
Protocluster galaxies having almost the same redshifts happen to be located near to each other in the spatial
dimension as well.
To discuss this more quantitatively, the histogram of the spatial separation from the nearest galaxy is shown in the
left panel of Figure \ref{D1st}.
The KS test suggests that the observed histogram is significantly different from a random distribution (the
p-value is less than 0.01); random distribution was generated from a uniform distribution in the limited-size
box of $17\times19\times29\,\mathrm{Mpc^3}$, which corresponds to the volume occupied by the ten member galaxies.
Furthermore, Figure \ref{D1st} shows a histogram of the separation from the nearest neighbor of field LAEs.
The data were taken from the spectroscopic LAE samples at $z=5.7$ and 6.5 \citep{kashikawa11}, which have a 
high degree of spectroscopic completeness ($\sim80-90\%$).
To compare protocluster and field galaxies more accurately, the difference in average number density between
protocluster and field are corrected.
The separation of field LAEs is multiplied by $(n_\mathrm{field}/n_\mathrm{protocluster})^{1/3}\sim0.6$, where
$n_\mathrm{protocluster}$ and $n_\mathrm{field}$ are the average number density of the protocluster and field,
respectively.
The p-value of a KS-test between the distribution of field LAEs and that of the protocluster was found to be
less than 0.03, suggesting that the excess of close pairs cannot be attributed to the clustering nature of
Ly$\alpha$ emitters alone.

\begin{figure}
\epsscale{1.1}
\plotone{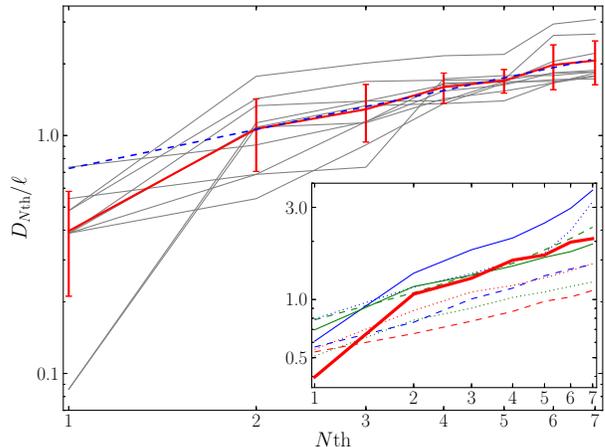}
\caption{Galaxy separation from the first to seventh nearest galaxies.
    In the vertical axis, the galaxy separation is normalized by $\ell$ ($=n^{-1/3}$, where $n$ is the number
    density).
    The red and gray lines represent the average and individual separations.
    The blue dashed line is the best fitted line to the separation from the second to seventh nearest galaxies.
    The inset shows the same relations in the case of protoclusters at lower redshifts.
    The solid lines shows the case for LBGs (red: SDF at $z=6.0$ (this study), green: SSA22 at $z=3.1$
    \citep{steidel98}, blue: MRC 0316-257 at $z=3.1$ \citep{kuiper12}).
    The dashed and dotted lines show the case for LAEs (dashed red: TN J1338-1942 at $z=4.1$, dashed green:
    TN J2009-3040 at $z=3.2$ \citep{venemans07}, dashed blue: MRC 0316-257 at $z=3.1$ \citep{venemans05}, dotted
    red: MRC 0943-242 at $z=2.9$, dotted green: MRC 0052-241 at $z=2.9$ \citep{venemans07}, dotted blue:
    MRC 1138-262 at $z=2.2$ \citep{pentericci00}).
    \label{DNth}}
\end{figure}

We also evaluated the separations from the second nearest galaxies (the right panel of Figure \ref{D1st}) based
on the same procedure as for the first nearest neighbors.
Its separation distribution was found to be reproduced by the random distribution with the p-value of 0.23.
Moreover, the separations from $N$th nearest galaxies were also calculated, and shown in Figure \ref{DNth}.
The separation, $D$, was fitted as a function of the $N$th ($N>2$) nearest galaxy: $\log(D)=a\times\log(N)+b$ ($a$
and $b$ are free parameters).
The values of $a$ and $b$ were $a=0.53\pm0.04$ and $b=1.82\pm0.07$ (blue dashed line in Figure \ref{DNth}).
The separation was found to be well approximated by this formula for $N>2$, while the observed separation from
the first nearest galaxies was found to be smaller with $1.8\sigma$ significance than the extrapolation from the
formula.
Even if we exclude two galaxies that have exceptionally small separations from the first nearest galaxy, the
trend was confirmed with $1.8\sigma$ significance.
Actually, as seen in Figure \ref{DNth}, all individual separations from the first nearest galaxy are smaller than
the best fitted line.
It is a statistically reliable result that galaxy separation from the first nearest galaxy is significantly smaller
in this protocluster.
Therefore, the protocluster galaxies tend to make galaxy pairs rather than triplets or larger structures.
However, the typical separation length ($>100\,\mathrm{kpc}$ in physical units) is too large for galaxy mergers or
interactions.
This is consistent with \citet{cooke10} who found that the fraction of Ly$\alpha$ emission in LBGs is larger in
pairs with separation of only $\leqslant70\,\mathrm{kpc}$ in physical units.
We could not find any strong correlation between the separation length and observed properties ($M_\mathrm{UV}$,
$L_\mathrm{Ly\alpha}$, and $EW_\mathrm{rest}$).

The inset in Figure \ref{DNth} shows the same relations in the case of protoclusters at lower redshifts taken from
the literature \citep{steidel98,venemans07,kuiper12}.
Most of the protoclusters, whose members have been identified by Ly$\alpha$ emission, show the smooth relation
without a bend at the first nearest neighbor.
Interestingly, the protoclusters, SSA22 and MRC 0316-257, whose members are selected by the dropout technique,
show a similar trend as our study: smaller separations from the first nearest galaxies than those from the
second or higher nearest galaxies.
Generally, LBGs have brighter UV luminosity than LAEs; thus, this trend would imply that bright galaxies are
located at the core of a protocluster and make pair-like structures, while faint galaxies are more widely
distributed.
Bright galaxies, presumably more massive, would form structures faster than less massive galaxies.
However, this comparison requires many caveats because the limiting magnitudes and spectroscopic completeness
are different; in most lower-redshift protoclusters, only $40-60\,\%$ galaxies are spectroscopically observed.
We only selected LBGs or LAEs in the above analysis, but various galaxy populations were found in $z\sim2-3$
protoclusters.
For example, \citet{kuiper11,kuiper12} found large subgroups in protoclusters at $z\sim2-3$, which contain
H$\alpha$ and [\ion{O}{3}] emitters as well.
These results would be consistent with the hierarchical structure formation model: at first, galaxies form small
groups like galaxy pairs, and these small groups grow to larger structures through mergers.
The cosmic epoch of $z\sim6$ may be the onset of cluster formation.

\subsection{Ancestors of Large-Scale Structure} \label{LSS}
\begin{figure}
\epsscale{1.1}
\plotone{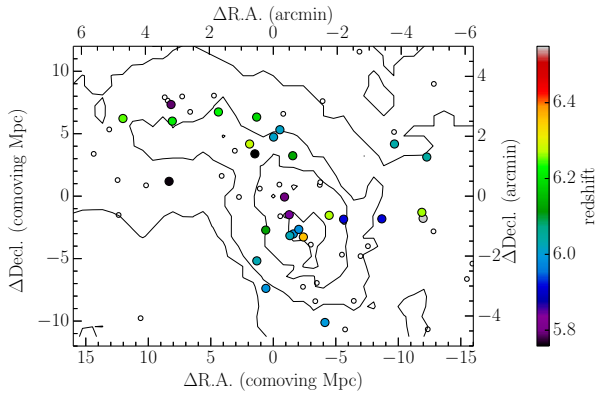}
\caption{Sky distribution of the Ly$\alpha$ detected galaxies (filled circles) and the $i'$-dropout galaxies
    (open circles).
    The redshifts of the Ly$\alpha$ detected galaxies are shown by the color code indicated in the legend at
    the right.
    The origin is the same as that of Figure \ref{sky}.
    The redshift of the protocluster corresponds to the light blue, and $z\sim6.2$ galaxies (light green) appear
    to align in the east-west direction at $\Delta\mathrm{Decl.}\sim6\,\mathrm{Mpc}$, (see the text in the
    discussion \S\ref{LSS}).
    \label{lss}}
\end{figure}

An interesting distribution of galaxies can be found behind the protocluster region.
Four galaxies at $z\sim6.2$ (color coded by light green in Figure \ref{lss}) seem to align in the east-west
direction at $\Delta\mathrm{Decl.}\sim6\,\mathrm{Mpc}$, like a filamentary structure, along the $2\sigma$
overdense region.
This filamentary structure, which is $80\,\mathrm{Mpc}$ away from the protocluster, is unlikely to merge with the
protocluster by $z=0$.
However, it is possible that these four galaxies may become another galaxy group by $z=0$ because they span only
$<6\,\mathrm{arcmin}$ and $\Delta z<0.06$. 
It should be noted that the length of large-scale filaments, as seen in the local universe, can reach
more than $100\,\mathrm{Mpc}$ \citep[e.g.,][]{park12,smargon12,alpaslan14}.
These galaxies would thus be part of the large-scale structure around a cluster at $z=0$.
Therefore, the protocluster at $z=6.01$ might be embedded in a large-scale overdense region.
\citet{einasto14} found that galaxy properties depend on the global environment like a supercluster, as well as
the local environment like a galaxy group.
Thus, it will be interesting to carry out further follow-up observations with infrared imaging on this protocluster
region in order to investigate galaxy properties in various environments, such as protoclusters, filaments, and
fields.

\section{CONCLUSIONS} \label{conc}

In this study, we have presented a systematic spectroscopic observation of the highest redshift protocluster
known to date.
We performed spectroscopy for all 53 $i'$-dropout galaxies down to $27.09\,\mathrm{mag}$ in $z'$-band in/around the
protocluster region, and Ly$\alpha$ emission of $EW_\mathrm{rest}\gtrsim11\,\mathrm{\AA}$ should be completely
detected with $\mathrm{S/N}>3$ in our spectroscopy.
The results and implications of this study are summarized below.

\begin{enumerate}
\item From this observation, 13 galaxies were newly identified as real $z\sim6$ galaxies by detecting their
Ly$\alpha$ emission lines.
Combined with the sample from \citet{toshikawa12}, 28 galaxies were confirmed to be located at $z\sim6$.
Ten of these are clustered in a narrow redshift range between $z=5.984$ and $z=6.047$
($\Delta z\lesssim0.06$), corresponding to a radial separation of $26.1\,\mathrm{Mpc}$.

\item We compared our results with the light-cone models, in which we applied the same $i'$-dropout selection
and the same overdensity measurements as in the observation.
We obtained a relation between the observed overdensity significance at $z\sim6$ and the predicted dark matter halo
mass at $z=0$.
Based on this relation, the observed protocluster with $6\sigma$ significance is predicted to grow into a galaxy
cluster ($\sim5\times10^{14}\,\mathrm{M_\sun}$) at $z=0$.

\item We also derive the spatial map of probability for a galaxy located at a certain position relative to the
protocluster center to become a cluster member at $z=0$.
Based on the map, the ten galaxies within $10\,\mathrm{Mpc}$ from the center of the protocluster are found, with
more than an 80\% probability, to merge into a single dark matter halo by $z=0$.
This clustering of galaxies is consistent with the progenitor of a galaxy cluster.

\item No significant difference in observed galaxy properties, such as $M_\mathrm{UV}$, $L_\mathrm{Ly\alpha}$,
and $EW_\mathrm{rest}$, between the protocluster and the field are found except for the brightest galaxies.
This suggests that this protocluster is still in the early phase of cluster formation before the onset of any
obvious environmental effects.
Further multi-wavelength imaging to trace their SEDs from the rest-frame optical to far-infrared wavelength will
provide us with clues about possible differences in stellar mass, dust, and gas mass. 

\item The velocity dispersion of this protocluster is too high ($869\pm85\,\mathrm{km\,s^{-1}}$) to consider
virial equilibrium.
This high velocity dispersion would be caused by the internal three-dimensional structure of the protocluster.
We also obtained a statistically reliable result that galaxies tend to form close pairs in the protocluster.
These pair-like subgroups will coalesce into a single halo and grow into more massive structures through accretion
of material from their surroundings.
Filamentary structure was found to be $80\,\mathrm{Mpc}$ away from the protocluster.
Although it is not expected to merge with the protocluster, this could become a large-scale structure around the
galaxy cluster, as seen in the local universe.

\end{enumerate}

As we have shown, the results obtained in this study are qualitatively and, in many ways, also quantitatively,
consistent with the hierarchical structure formation model.
The epoch of $z\sim6$ may be the time when galaxies start to coalesce in order to form a cluster as seen in the
local universe. 
It is necessary to increase the number of observations of protoclusters at $z\sim6$ and even at lower redshifts
to obtain statistical samples of protoclusters across cosmic time, in order to clarify the general features of
cluster formation and galaxy evolution in dense regions.
Using the new instrument Hyper SuprimeCam (HSC) on the Subaru telescope, we are carrying out an unprecedentedly
wide and deep survey over the next five years.
Based on the number density estimate of protoclusters in Section \ref{model}, $>10$ $z\sim6$, $>500$ $z\sim5-4$
protoclusters, and even one $z\sim7$ protocluster would be discovered from this Subaru strategic survey.
These will enable us to derive a more complete picture of cluster formation and galaxy evolution in high density
environments.

\acknowledgments
Some of the data presented here were obtained at the W.M. Keck Observatory, which is operated as a scientific
partnership among the California Institute of Technology, the University of California, and the National
Aeronautics and Space Administration. 
The Observatory was made possible by the generous financial support of the W.M. Keck Foundation.
The other data were collected with the Subaru Telescope, which is operated by the National Astronomical
Observatory of Japan.
We are grateful to the W.M. Keck and Subaru Observatory staff for their help with the observations.
We are also grateful to Dr. Takashi Hattori, Subaru Sr. Support Astronomer, who helped us reduce the spectroscopic
data obtained by Keck/DEIMOS.
The Millennium Simulation databases used in this paper and the web application providing online access to them
were constructed as part of the activities of the German Astrophysical Virtual Observatory (GAVO).
We thank the anonymous referee for valuable comments and suggestions which improved the manuscript.
This research was supported by the Japan Society for the Promotion of Science through Grant-in-Aid for Scientific
Research 23340050 and 12J01607.

{\it Facilities:} \facility{Subaru (FOCAS)}, \facility{Keck:II (DEIMOS)}

\end{document}